\documentclass[prl,twocolumn,superscriptaddress,showpacs]{revtex4-1}
\usepackage{amsmath}
\usepackage{amssymb}
\usepackage{xcolor}
\usepackage{hyperref}
\hypersetup{colorlinks=true,citecolor={blue},linkcolor={blue},urlcolor={blue}}
\usepackage{graphicx}
\usepackage{dcolumn}
\newcommand{\mi}{\mathrm{i}}
\newcommand{\ve}{\varepsilon}
\newcommand{\half}{\textstyle{\frac{1}{2}}}
\newcommand{\ihalf}{\textstyle{\frac{\mathrm{i}}{2}}}
\newcommand{\quart}{\textstyle{\frac{1}{4}}}

\begin{document}

\title{Spontaneous polariton currents in periodic lateral chains}
\author{A. V. Nalitov}
\affiliation{School of Physics and Astronomy, University of Southampton, Southampton SO17 1BJ, United Kingdom}
\author{T. C. H. Liew}
\affiliation{Division of Physics and Applied Physics, School of Physical and Mathematical Sciences,
Nanyang Technological University, Singapore 637371}
\author{A. V. Kavokin}
\affiliation{School of Physics and Astronomy, University of Southampton, Southampton SO17 1BJ,
United Kingdom}
\affiliation{CNR-SPIN, Viale del Politecnico 1, I-00133, Rome, Italy}
\affiliation{Spin Optics Laboratory, St. Petersburg State University, St. Petersburg, 198504, Russia}
\affiliation{Russian Quantum Center, 100 Novaya Street, Skolkovo, Moscow Region 143025, Russia}
\author{B. L. Altshuler}
\affiliation{Physics Department, Columbia University, New York, New York 10027, USA}
\author{Y. G. Rubo}
\affiliation{Instituto de Energ\'{\i}as Renovables, Universidad Nacional Aut\'onoma de M\'exico, Temixco, Morelos 62580, Mexico}
\affiliation{Center for Theoretical Physics of Complex Systems, Institute for Basic Science (IBS), Daejeon 34051, Republic of Korea}

\begin{abstract}
We predict spontaneous generation of superfluid polariton currents in planar microcavities with lateral periodic modulation of both potential and decay rate. A spontaneous breaking of spatial inversion symmetry of a polariton condensate emerges at a critical pumping, and the current direction is stochastically chosen. We analyse the stability of the current with respect to the fluctuations of the condensate. A peculiar spatial current domain structure emerges, where the current direction is switched at the domain walls, and the characteristic domain size and lifetime scale with the pumping power.
\end{abstract}


\maketitle

Being weakly interacting composite bosons, exciton-polaritons undergo Bose-Einstein condensation (BEC) \cite{Balili2007,Kasprzak2006} and  may exhibit superfluid behaviour \cite{Amo2009,Amo2009a,Sanvitto2010}.
Its striking manifestation: persistent, frictionless polariton currents may be used for information exchange between optical logical devices \cite{Liew2008,Liew2010}.
Like atomic or molecular counterparts, polariton superfluids also sustain
quantized vortices \cite{Lagoudakis2008,Sanvitto2010} and half-vortices
\cite{Rubo2007,Lagoudakis2009,Hivet2012}.
Polariton transport free of backscattering may be as well realised in the linear regime at the edges of polariton topological insulators
\cite{Nalitov2015a,Karzig2015,Bardyn2015}.
In all cases strong light-matter coupling plays a crucial role, as it supplements light cavity photons with strong nonlinearity and
provides strong magneto-optical interaction.

Contrary to the cold atom systems, cavity polaritons are characterized by finite lifetime, limited by the photon escape from the cavity.
Formed as a result of the compensation of this dissipation by continuous pumping from  exciton reservoirs, the polariton condensates are thus out of thermal equilibrium.
In the case of nonresonant optical or electric pumping the quantum coherence of the condensate is formed spontaneously.
For sufficiently fast polariton thermalisation the condensate is formed in the single particle ground state in full analogy
with the BEC.
However, polaritons with slow energy relaxation can choose an exited single-particle state for macroscopic occupation \cite{Lai2007a,Kim2011,Winkler2016}. 
Such a state is usually degenerate.
Moreover, when single-particle states possess different lifetimes,
the interaction between polaritons can lead to condensation into specific many-particle states with spontaneously broken symmetries such as the time-reversal and parity symmetries \cite{Aleiner2012}.
In this weak lasing state, the system is stabilized by the repulsive
polariton-polariton interactions rather than the reservoir depletion (gain-saturation nonlinearity).
The combined effect of interactions and gain-saturation extends the stability of weak lasing states to high excitation powers \cite{Ohadi2015}.
These states can be easily manipulated and switched experimentally \cite{Ohadi2016a,Dreismann2016}.

Polariton condensates are commonly described in the mean field approximation with non-Hermitian Hamiltonians accounting for both decay and external pumping.
In this Letter we consider a lateral periodic complex potential for polaritons in planar microcavities or microwires similar to
those realized in Ref.\ \cite{Lai2007a,Zhang2015,Gao2016a}.
Its imaginary part, corresponding to spatially dependent polariton decay rate, is determined by the spatial modulation of the cavity quality factor.
In turn, the real part of the potential may be realized with spatial quantization energy modulation of either the photonic or excitonic component.
Regardless of realization of this potential the single-polariton mode with longest life-time turns out to be at the edge of the lowest energy mini gap.
Assuming the feed for all Bloch wave modes close to the bottom of the polariton dispersion being equal, this mode has the lowest lasing threshold.

In the case of in-phase modulation of the real and imaginary parts of the potential, the second threshold emerges due to the repulsive interaction.
While below this threshold the condensate order parameter period coincides with the modulation lattice constant, crossing it results in abrupt period doubling.
There are two degenerate double period condensate states connected by the lattice translation spontaneous symmetry breaking.
In terms of the two mode approximation the transition is described as an admixture of the ground polariton state having an intermediate lifetime to the macroscopically occupied second band bottom state having the longest lifetime
\citep{Zhang2015}.

What happens if the modulations of the real and imaginary parts of the potential have opposite phases?
In this case the lowest threshold corresponds to the polariton mode at the top of the lowest miniband rather than at the bottom of the second miniband.
The repulsive interaction blueshifts the initial condensate, causing the dynamical admixture of the second band bottom state.
As we show below, there appears a nontrivial (neither $0$ nor $\pi$) phase difference between these two states constituting the condensate, which manifests itself in a nonzero net polariton current.

It should be noted, that apart from some similarity, the formation of spontaneous current is different in origin from the Kibble-Zurek mechanism of vorticity \cite{Zurek1985}.
The polariton condensation does not follow the standard Kibble-Zurek scenario \cite{Matuszewski2014,Solnyshkov2016b}, and the current does not appear because of nonadiabaticity of transition.
While the nonzero current is general feature of nonlinear Bloch solutions \cite{Chestnov2016}, in this Letter we show that in the present case there is no stationary condensate state without the flux, and this flux is not quantized.

\begin{figure} [t]
\includegraphics[width=0.99\columnwidth]{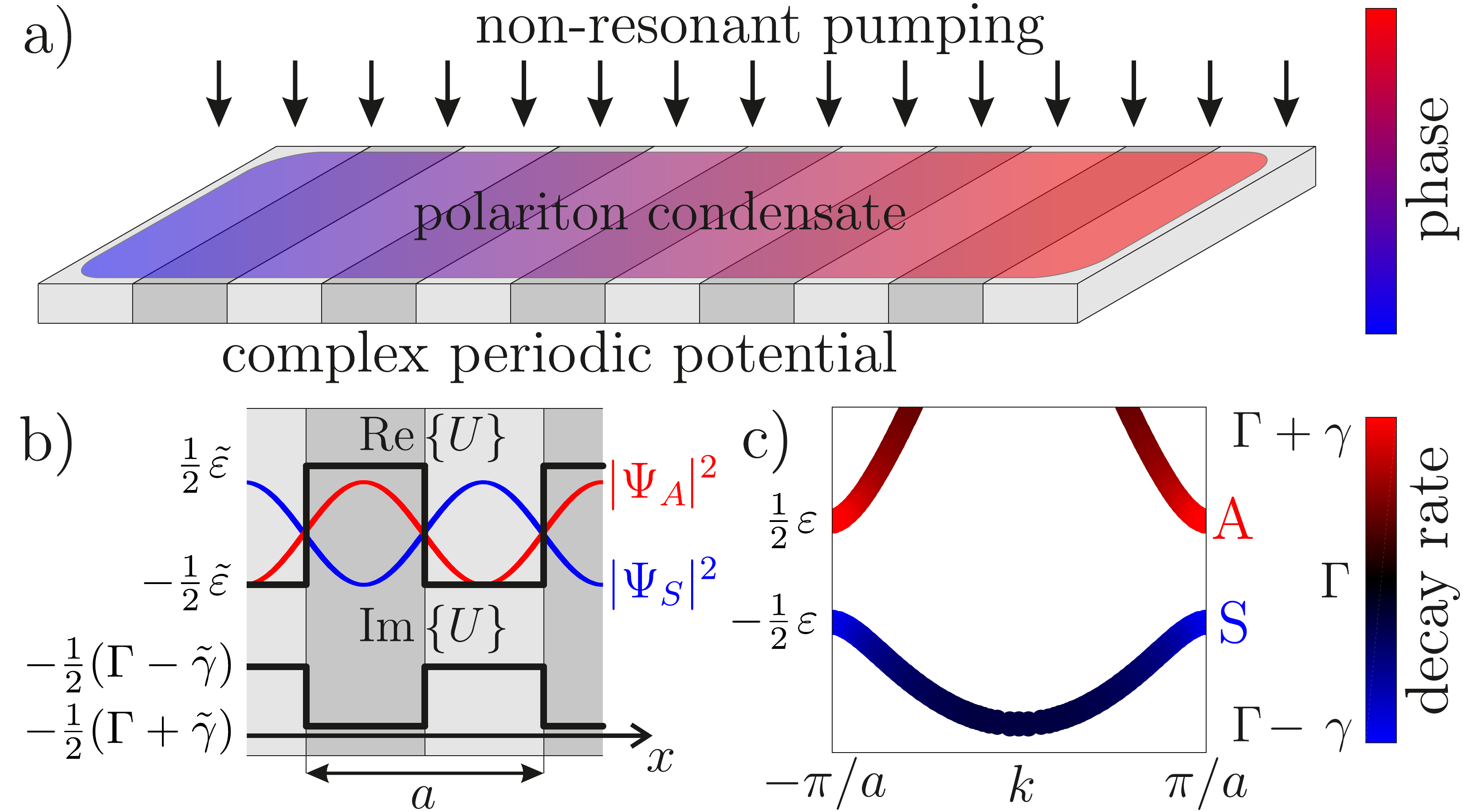}
 \caption{\label{fig_1} a) Sketch of the proposal: polariton current of spontaneously chosen
 direction flows in a condensate fed by a symmetric homogenous pumping.
 b) Spatial distribution of real and imaginary parts of the periodic potential $U(x)$ and
 distribution of the first and second band wavefunctions at the Brillouin zone edge.
 c) Complex polariton band structure in the Kronig-Penney model. The decay rate is shown with color.
 The fastest(slowest) decay rate characterises A(S) mode.}
\end{figure}

We solve the Gross-Pitaevskii equation (GPE) for the condensate wavefunction taking into account two types of nonlinearities, stemming from polariton repulsion and
reservoir depletion ($\hbar=1$):
\begin{equation}
 \label{eq_mainGP}
 \mi\frac{\partial\Psi}{\partial t}
 = \left[ - \frac{1}{2m}\frac{\partial^2}{\partial x^2} + U
 + \frac{\mi}{2}\left( W - \frac{\eta}{2}\langle\vert\Psi\vert^2\rangle \right)
 + \alpha \vert \Psi \vert^2 \right] \Psi.
\end{equation}
Here $m$ is the polariton effective mass and $\alpha$ is the interaction constant.
The pumping power $W$, detemined by reservoir population, is locally reduced due to its depletion, which is proportional to the condensate density averaged over the unit cell with the prefactor $\eta/2$.
This nonlinearity can be obtained as a result of exclusion of the equation for reservoir density \cite{Keeling2008}.
The averaging describes washing out the spatial inhomogeneities of the reservoir density by the exciton diffusion and excludes unphysical solutions with periodic modulation of the reservoir density.
We consider a lateral complex periodic potential $U(x)$ for polaritons, shown in Fig.\
\ref{fig_1}(a),
\begin{equation}
 \label{eq_U}
 U(x) = \left\lbrace \begin{matrix}
 -\half[\tilde{\ve} + \mi(\Gamma-\tilde{\gamma})], & \vert x - na \vert < \quart a, \\[.5em]
 \half[\tilde{\ve} - \mi(\Gamma+\tilde{\gamma})],  & \quart a < \vert x -na \vert < \half a,
\end{matrix} \right.
\end{equation}
where $n$ spans all integers.

The Kronig-Penney model \cite{Bastard1988} for a polariton in such a potential yields the band structure (Fig.\
\ref{fig_1}(b)), where the longest (shortest) lifetime characterizes the lowest-band top (the
second-band bottom) state, which we denote as S(A). In the nearly free particle approximation, the
two polariton modes $\Psi_S\propto\cos(k_0 x)$ and $\Psi_A\propto\sin(k_0x)$ with $k_0=\pi/a$ are
separated by the energy band gap $\ve=2\tilde{\ve}/\pi$ and decay at rates $\Gamma-\gamma$ and
$\Gamma+\gamma$, respectively, where $\gamma=2\tilde{\gamma}/\pi$.

Assuming slow polariton thermalization and equal feeding of the modes from the reservoir one should expect the longest lifetime mode (S) to cross the
lasing threshold first with increasing pumping power.
With further growth of the condensate population the repulsive interaction blueshifts the condensate and eventually leads to admixture of the second band bottom (A) state.
We project Eq.\
\eqref{eq_mainGP} onto the plane-wave two-mode basis and search for the solution in the form $\Psi =
\psi_+\exp(+\mi k_0x)+\psi_-\exp(-\mi k_0x)$.
Assuming that the envelopes $\psi_\pm$ are smooth on the scale of the lattice parameter $a$ we neglect second spatial derivatives of $\psi_\pm$ and obtain
\begin{equation} \label{eq_2modes}
 \left[\frac{\partial}{\partial t} \pm c\frac{\partial}{\partial x} +
 \frac{g(s)}{2} + \frac{\mi\alpha}{2} \left( 3s \mp s_z \right) \right] \psi_\pm
 = \frac{\gamma+\mi\ve}{2}\psi_\mp.
\end{equation}
Here $c=\pi/ma$, $g(s) = \eta s - w$ with $w = W-\Gamma$, $s=(|\psi_+|^2+|\psi_-|^2)/2$, and
$s_z=(|\psi_+|^2-|\psi_-|^2)/2$.
In what follows, we first find spatially homogeneous solutions
\begin{equation}
 \label{eq_psipm}
  \psi_\pm(x,t)\equiv\psi_\pm(t)=\sqrt{s\pm s_z}\,e^{-\mi(\Omega t\pm\phi)},
\end{equation}
with time-independent $s$, $s_z$, the phase shift $\phi$, and the emission frequency $\Omega$
(counted from the middle of the gap).
Then we study their stability with respect to small spatially nonhomogeneous fluctuations.
See the Supplemental Material for details of the derivations.

The first pair of solutions preserve the parity symmetry of Eqs.\ \eqref{eq_2modes}, i.e. $s_z=0$, which corresponds to the condensation in the single-particle S state with $s=(w+\gamma)/\eta$ and $\phi=0$, and to the condensation in the A state with $s=(w-\gamma)/\eta$ and $\phi=\pi/2$.
As expected, the S mode solution has the lowest
threshold pumping power $w = -\gamma$.

For the second pair of stationary solutions characterized by nonzero $s_z$ and nontrivial phase difference $\phi$,
\begin{equation}
 \label{eq_szphi}
 s_z = -\frac{\ve}{\gamma} s \tan(2\phi), \quad
 \tan(2\phi)=\pm\sqrt{\frac{\gamma^2-g(s)^2}{\ve^2+g(s)^2}},
\end{equation}
we obtain for the population $s$ and emission frequency $\Omega$:
\begin{equation}
 \label{eq_s_vs_w}
 \ve\alpha s g(s) = \gamma \left[\ve^2 + g(s)^2\right], \quad
 \Omega = \frac{1}{2}\left[ 3 \alpha s - \frac{\ve\gamma}{g(s)} \right].
\end{equation}

The first equation in \eqref{eq_s_vs_w} has two roots for the condensate population $s$. However,
complemented with restrictions $s>0$ and $0<g(s)<\gamma$, it has two branches of solutions for
pumping powers above the critical point 
\begin{equation}
w_c = \eta {\gamma^2+\ve^2 \over \ve\alpha}-\gamma,
\end{equation}
if $\alpha/\eta<(\gamma/\ve)-(\ve/\gamma)$ and a single branch of solutions otherwise.
These two regimes correspond to subcritical and supercritical pitchfork bifurcations, respectively \cite{Strogatz1994}.
We refer to them as to transitions of types I and II respectively in analogy to phase transitions of the first and the second order.
Note that the condition $0<g(s)<\gamma$ implies that these solutions
correspond to intermediate condensate populations between populations of the symmetric S and antisymmetric A solutions.

\begin{figure}
\includegraphics[scale=0.183]{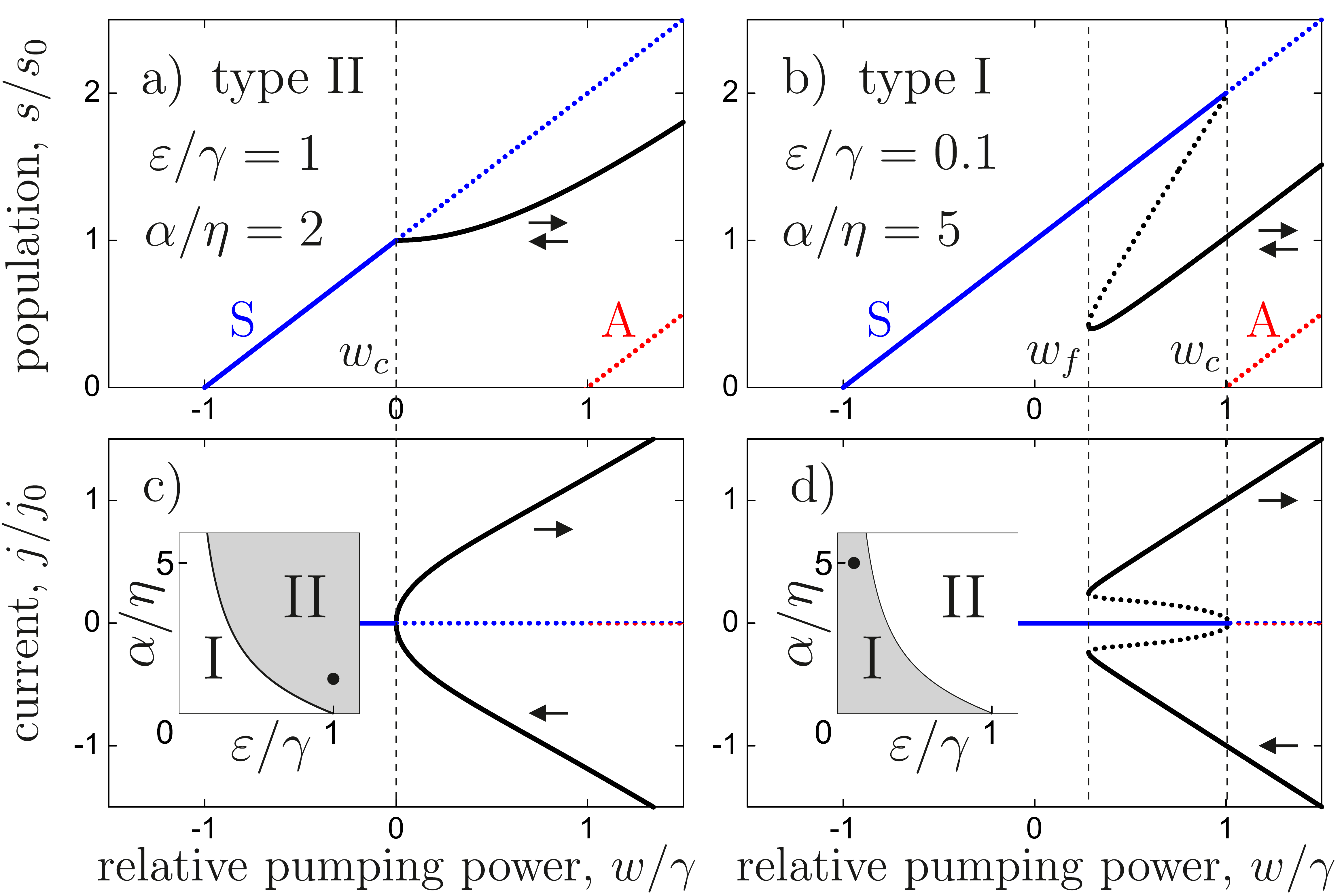}
 \caption{\label{fig_2} (a,b) Condensate population dependence on the pumping power.
 (c,d) Imbalance of populations $s_z$ defining the value of the net polariton current.
 Uniformly stable(unstable) stationary solutions are plotted with solid(dotted) lines.
 Blue and red lines correspond to symmetry conserving condensates at S and A polariton modes.
 The two sets of parameters correspond to the presence (type I) and absence (type II) of a tristability
 region between the S state solution and the pair of symmetry breaking solutions.
 In the former case two hysteresis loops emerge.
 The lower two plots are zoomed to the pitchfork and saddle-node bifurcation points $w_c$ and $w_f$.
 Inset: phase transition type depending on $\alpha/\eta$ and $\gamma/\varepsilon$. }
\end{figure}

The branches of fixed points for both cases are shown in Fig.\ \ref{fig_2}, $s(w)$ shows the normalized condensate population, while the nonzero population imbalance $s_z(w)$ defines the polariton current flowing in the condensate $j = 2k_0s_z/m$.
For A and S state solutions, shown in red and blue respectively, there is no current and populations are linear in the pumping power.
The nonzero current is a property of the other pair of branches, plotted in black.
The two solutions have the same population, but the current directions are opposite.
It is important to note that the polariton density for the current solutions is lower than the maximal possible density achieved at the unstable state S.
This is characteristic for the weak lasing regime, where the losses in the polariton system are adjusted to compensate the gain rather than minimized.

Fig.\ \ref{fig_2} also indicates the stability of the solutions with respect to homogeneous fluctuations, the unstable branches are shown with dotted lines.
While the A state is always unstable, the S state solution is stable at $s<s_c=(\ve^2+\gamma^2)/\alpha\ve$, below the critical pumping value $w_c$. The symmetry breaking solutions, on the contrary, are stable above the critical pumping in the case of type II bifurcation.
For type I with two nontrivial solutions for $s(w)$, the lower branch is
stable, while the upper one is unstable.
The critical point $w = w_c$ is thus a pitchfork bifurcation of a cusp catastrophe.
It is supercritical for type II and subcritical for type I.
The latter is accompanied with a region of tristability below the critical point.
It is limited from below with a saddle-node bifurcation at $w=w_f=2\sqrt{\gamma\eta(\gamma\eta-\ve\alpha)}/\alpha < w_c$.
The tristability region corresponds to hysteresis loops between the symmetric S state solution and the pair of symmetry breaking solutions.
The sign of $s_z$ and thus the direction of the polariton current $j$ is
spontaneously chosen by the system once the condensate passes the critical point.
Its value adiabatically grows from zero in the supercritical case and appears abruptly in the the subcritical one.
On the way back along a randomly chosen hysteresis loop the current abruptly disappears at the saddle-node bifurcation point $w = w_c$.

\begin{figure}[t]
\includegraphics[scale=0.4]{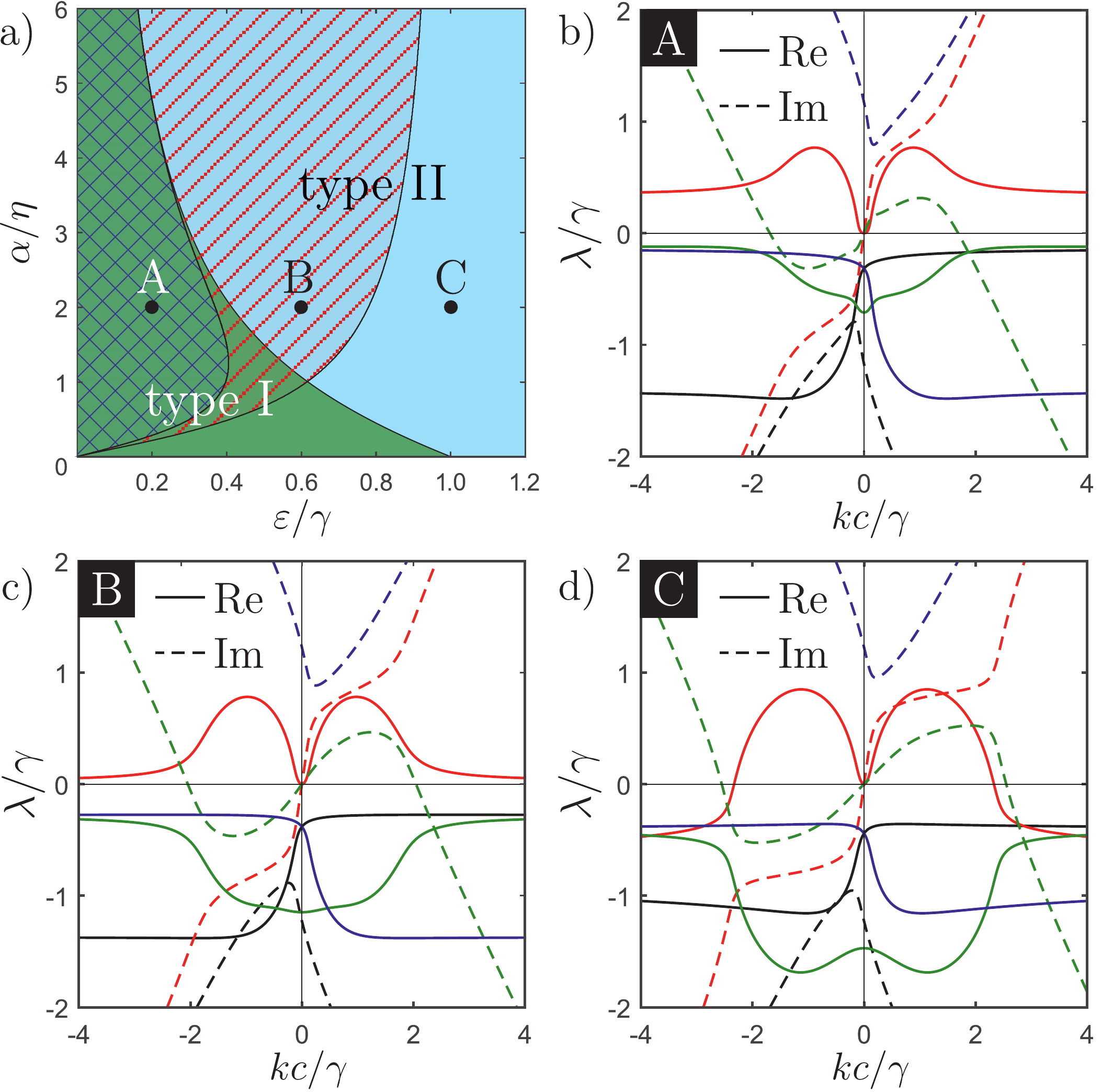}
 \caption{\label{fig_3}
 (a) Stability diagram. Type I and II transitions are shown with green and blue colours respectively.
 Regimes of current stability with respect to short wavelength fluctuations are shown with hatching:
 the blue double hatched area corresponds to the unstable regime;
 the red single hatched area covers the parameter space where the current is stable in a range of
 pumping powers; the area free of hatching corresponds to the stable current regime.
 (b-d) The Lyapunov spectra of condensate current solutions. Solid and dashed lines show the real
 and imaginary parts, respectively. Parameters correspond to the black dots in the panel (a).
 The pumping power is slightly above the critical point.}
\end{figure}

To study the stability with respect to inhomogeneous phase and population fluctuations, we calculate the elementary excitation spectra.
In the standard way, we linearize Eq.\
\eqref{eq_2modes} with respect to plane wave perturbation
\begin{equation}
 \label{eq_perturb}
 \delta\psi_\pm(x,t) = e^{-\mi\Omega t}\left(u_\pm e^{\mi kx+\lambda t} + v^*_\pm e^{-\mi kx+\lambda^* t}\right)
\end{equation}
of the spatially uniform solutions described above.

The dispersion of the real and imaginary parts of the Lyapunov exponent $\lambda(k)$ for the nonzero current solution is plotted in Fig.\ \ref{fig_3} for different regimes.
Here we consider uniformly stable solutions, characterized by three modes with $\mathrm{Re}\lambda(0)<0$ and one Goldstone mode with $\lambda(0)=0$.
The Goldstone mode appears due to the irrelevance to the global shift of the total phase of the condensate $\Phi$.

The short wavelength limit analysis provides an important condition of the applicability of the two-mode approximation \eqref{eq_2modes}.
Depending on the parameters, there are three regimes, illustrated by Fig.\ \ref{fig_3}(a).
For a given nonlinearity relation $\alpha/\eta$, low values of
$\ve/\gamma$ correspond to the instability of the symmetry breaking condensate in the short wavelength limit.
In this domain, the system should exhibit either period doubling bifurcations \cite{Zhang2015} or strongly chaotic behavior.
There is an intermediate regime where the spontaneous current condensate is stable in a certain range of pumping powers.
However, in the most realistic case of large $\ve/\gamma$, the spontaneous current solutions are stable with respect to short wavelength fluctuations and are well described by Eqs.\ \eqref{eq_2modes}.
Note that for the long wavelength fluctuations the condition $ka\ll1$ is well satisfied.
The two parameters defining the short wavelength stability, $\alpha/\eta$ and $\ve/\gamma$ depend on the system design: the interaction nonlinearity $\alpha$ may be controlled by polariton lateral confinement (etched microcavity width), while $\ve$ and $\gamma$ are independently tunable through periodic modulation of the cavity photonic mode energy and broadening.

In the case that the condensate is stable with respect to short wavelength fluctuations there is still a region of positive Lyapunov exponents, as it is seen from Fig.\ \ref{fig_3}(d).
We note that this result does not follow from Mermin-Wagner theorem \cite{Hohenberg1967}, although the latter also forbids long-range order in 1D.
A condensate in a long enough microcavity chain thus falls apart and is expected to transform into a polariton current domain structure.
The characteristic domain size $l_d\sim c/w$, deduced from the extremum
position of the excitation spectrum imaginary part, as well as its characteristic lifetime $\tau_d\sim 1/ w$, obtained as the inverse extremum value, scales as the inverse occupation number of the condensate.
Hence, the domain wall characteristic speed $v_d \sim l_d/\tau_d \approx c$ is independent on the pumping power.

On the other hand, a finite system with periodic boundary conditions, such as a microcavity ring chain, may support a global bifurcation towards a polariton condensate with spontaneously chosen and persistent circular current.
This is possible in the case of short wavelength stability in a range of low pumping powers.
The upper boundary of this range is determined by the cut-off fluctuation wave vector $k_c$ defined by $\mathrm{Re}\lambda(k_c)=0$ [see
red curve in Fig.\ \ref{fig_3}(d)].
The persistent current is possible in the ring with radius $R<k_c^{-1}$.
We note that this mechanism of formation of stable polariton currents in modulated polariton rings is different to that in unmodulated polariton rings, in which high angular momentum states were shown to be unstable \cite{Li2015}.
Here the circular polariton current is formed spontaneously and is not inherited from an optical pump.

\begin{figure}[t]
\includegraphics[scale=0.32]{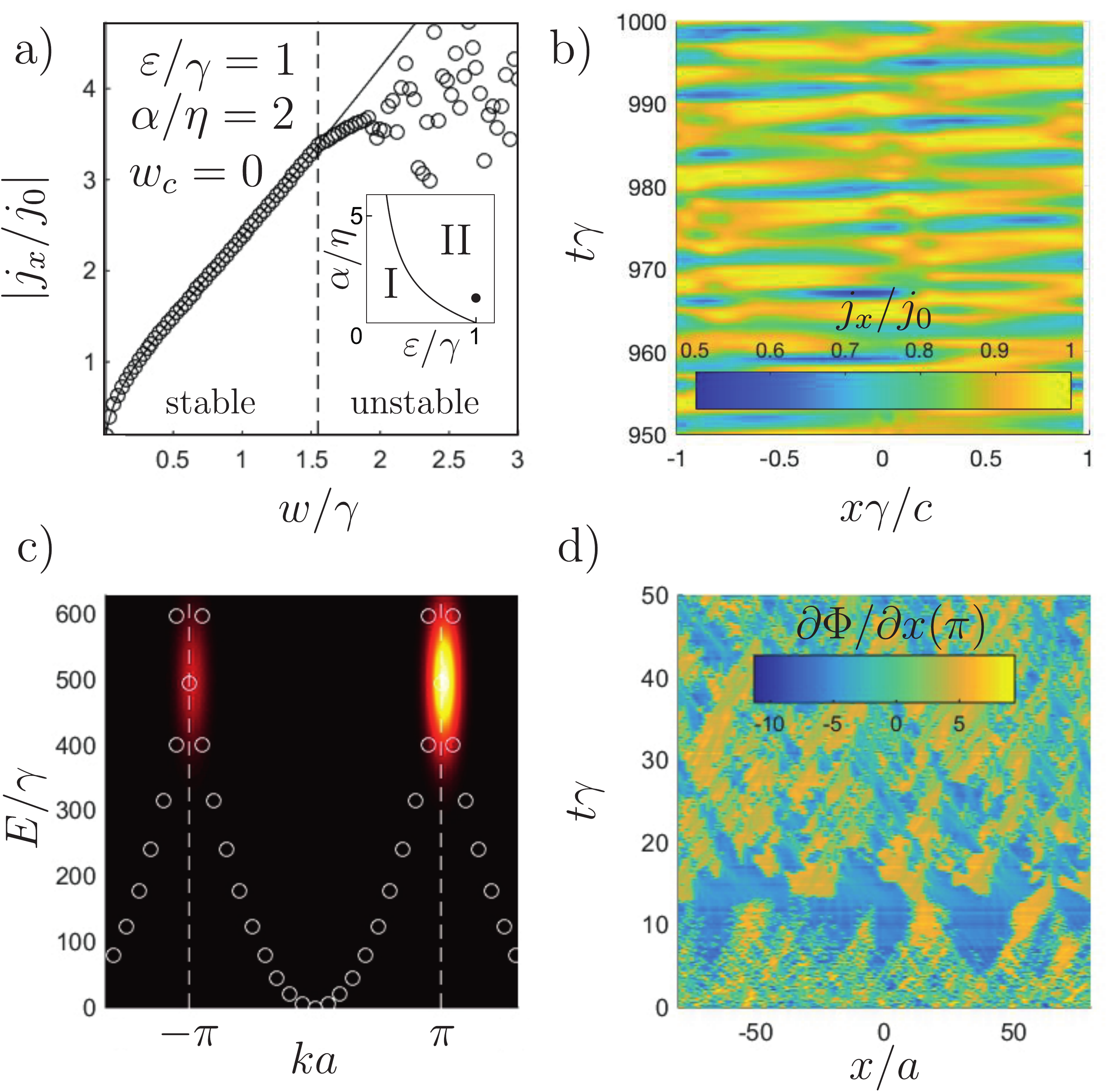}
 \caption{\label{fig_4}
 (a,b) Numerical simulation in the two-mode approximation.
 (a) Absolute value of the spatially averaged current vs pumping power. The system finite size
 defines the instability onset point. Parameters correspond to the type II transition with the
 finite system size set by $-c\leq x\gamma\leq c$.
 (b) Spatial and temporal dependence of polariton current in the unstable regime
 (for $w=2.1\gamma$): domain structure emerges in an initially homogeneous condensate. \\
 (c,d) Full numerical simulation of GPE with periodic complex potential.
 (c) Condensate dispersion demonstrating spatial asymmetry. The intensity has been broadened in
 energy and momentum for visualization. Circles denote the bare dispersion, quantized by the finite
 system size. Parameters: $w=\gamma$, $-a\leq x\leq a$. d) Spatial and temporal dependence of local
 phase gradient in the unstable regime (given by larger system size $-80a\leq x\leq 80 a$),
 with chaotic evolution of domain structure.}
\end{figure}

Eqs.\ \eqref{eq_2modes} can be solved numerically, by propagating in time from an initial random noisy state.
The results of this numerics for the type-II transition are shown in Fig.~\ref{fig_4}(a).
A stable current is formed spontaneously from the initial noise
when the pump intensity is below the critical value set by the finite system size.
Above the critical pump intensity the system breaks up forming oscillating domains (see Fig.~\ref{fig_4}(b)).
For very large pump intensities a chaotic state forms with strong fluctuations in the spatially averaged current.

The spontaneous currents can also be obtained by a direct numerical solution of the original GPE \eqref{eq_mainGP}, without the two-mode approximation \eqref{eq_2modes}.
Here we add a Langevin noise term~\cite{Winkler2016}, which serves both as an initial seed for the condensate and a test of its stability to fluctuations.
Using type II parameters, we find preferential condensation in the $k=+k_0$ state (see Fig.~\ref{fig_4}(c)).
This state is further characterized by a relatively small and spatially uniform phase gradient (not shown), which is stable in time despite the presence of noise.
Repeating the calculation revealed random selection of the $\pm k_0$ states, with equal probability.
Finally, by increasing the system size, the spontaneous currents become unstable leading to the formation of chaotic domains, characterized by different phase gradients, which evolve spatiotemporally.

In summary, we considered polariton condensation in microcavities with potential and decay rate periodically modulated in space.
Our analysis suggests that such systems undergo a spontaneous symmetry
breaking and the formation of polariton currents.
We identify the critical conditions for this effect to emerge and produce the phase diagram showing type I and type II transition boundaries.
For large systems oscillating domains of counterpropagating currents are
predicted.
For systems smaller than the characteristic domain size, e.g., polariton rings, the spontaneously formed currents are stable and survive in the presence of spatiotemporal noise.
See the Supplemental Material for discussion of the experimental realization of the effect.

This work has been supported in part by CONACYT (Mexico) Grant No.\ 251808, by IBS-R024-D1, and the MOE (Singapore) grant 2015-T2-1-055.
AVK acknowledges the support from the Russian Foundation for Basic Research grant 15-59-30406, DFG ICRC project TR 160, and the EPSRC Established Career Fellowship in Quantum Polaritonics.

\bibliography{spontaneous5}

\widetext

\section{Supplemental material}

\section{Two modes model}
The Gross-Pitaevskii equation for polaritons in a periodic potential ($\hbar = 1$)
\begin{equation} \label{eq_GPE_full}
\mi {\partial \Psi \over \partial t} = \left[ -{1 \over 2 m} {\partial ^2 \over \partial x^2} + U(x) + {\alpha \over 2} \vert \Psi \vert^2 + {\mi \over 2} \left(W - {\eta \over 2} \langle \vert \Psi \vert^2 \rangle \right) \right] \Psi.
\end{equation}
Here $m$ is the effective polariton mass, $W$ is the external pumping term, stemming from exciton scattering from the reservoir towards the condensate, $\alpha>0$ and $\eta>0$ are nonlinearity prefactors stemming from polariton-polariton repulsive interaction and reservoir depletion, respectively.

The complex periodic potential, accounting for spatially modulated decay rate,
\begin{equation}
 U(x) = \left\lbrace \begin{matrix}
 -\half[\tilde{\ve} + \mi(\Gamma-\tilde{\gamma})], & \vert x - la \vert < \quart a, \\[.5em]
 \half[\tilde{\ve} - \mi(\Gamma+\tilde{\gamma})],  & \quart a < \vert x -la \vert < \half a,
\end{matrix} \right.
\end{equation}
where $l$ spans all integer, may be conveniently represented as a Fourier series
\begin{align}
U(x) &= \sum_{n=0}^{+ \infty} U_n \cos (2 \pi n x/a), \\
U_0 &= -\ihalf \Gamma, \nonumber \\
U_n &= -{2 \over \pi n}(\tilde{\ve}-\mi\tilde{\gamma})\sin\left(\pi n \over 2\right), \quad n=1,2,3,\dots . \nonumber 
\end{align}

The nearly free polariton approximation, valid in the limit $k_0^2/(2m) \gg \vert U_1 \vert$, where $k_0 = \pi /a$, yields complex energies $E_{S(A)} = k_0^2 / (2 m) + U_0 \pm U_1 / 2$ for the states at the edges of the first minigap:
\begin{equation}
E_S =  - {\mi \over 2} \Gamma - {1 \over \pi} \left( \tilde{\ve} - \mi \tilde{\gamma} \right), \qquad
E_A =  - {\mi \over 2} \Gamma + {1 \over \pi} \left( \tilde{\ve} - \mi \tilde{\gamma} \right).
\end{equation}
The band gap value is therefore $\ve \equiv 2 \tilde{\ve}/\pi$, the decay rates of the states S and A are $\Gamma - \gamma$ and $\Gamma + \gamma$ respectively, where $\gamma \equiv 2 \tilde{\gamma}/\pi$. Corresponding wavefunctions are
\begin{equation}
\Psi_\mathrm{S}\propto\cos(k_0 x), \qquad \Psi_\mathrm{A}\propto\sin(k_0 x).
\end{equation}

As the lowest threshold state is at one of the first minigap edges, we search for a solution of Eq.\ \eqref{eq_GPE_full} as a superposition of the two states. Namely, we substitute $\Psi = \psi_+ \exp(\mi k_0 x) + \psi_- \exp(-\mi k_0 x)$ into Eq.\ \eqref{eq_GPE_full} and neglect the second derivatives due to presumed smoothness of the envelopes $\psi_\pm$. This gives
\begin{equation} \label{eq_GP_2modes}
  \frac{\partial\psi_{\pm}}{\partial t} \pm c \frac{\partial\psi_{\pm}}{\partial x} 
  =-\frac{1}{2}\left[\Gamma - W + \frac{\eta}{2}(\vert\psi_+\vert^2 + \vert\psi_-\vert^2) 
                     +\mi\alpha(\vert\psi_\pm\vert^2 + 2\vert\psi_\mp\vert^2)\right]\psi_\pm 
   +\frac{1}{2}(\gamma +\mi\ve)\psi_\mp,
\end{equation}
where $c = k_0/m$ and the energy is counted from the center of the first minigap.

\section{Pseudospin evolution equation}

We make a change of variables in Eq.\ \eqref{eq_GP_2modes}, introducing the pseudospin $\mathbf{s}$ with the components
\begin{equation} \label{def_spin}
s_x = \mathrm{Re} \left\lbrace \psi_+^*\psi_- \right\rbrace, \qquad
s_y = \mathrm{Im} \left\lbrace \psi_+^*\psi_- \right\rbrace, \qquad
s_z = (\vert \psi_+ \vert^2 - \vert \psi_- \vert^2)/2.
\end{equation}
Note that the definitions \eqref{def_spin} may be as well written as
\begin{equation}
\vert \psi_\pm \vert^2 = s \pm s_z, \qquad \psi_\pm^* \psi_\mp = s_x \pm\mi s_y.
\end{equation}

Multiplication of Eq.\ (\ref{eq_GP_2modes}) by $\psi_\pm^*$ yields
\begin{equation} \label{eq_psi1}
\left( {\partial \over \partial t} \pm c {\partial \over \partial x} \right) \vert \psi_\pm \vert^2 =
-\left( \Gamma - W + \eta s \right)\vert \psi_\pm \vert^2 + \gamma s_x \mp \varepsilon s_y,
\end{equation}
or, in terms of the pseudospin and effective decay rate $g(s) = \Gamma - W + \eta s$:
\begin{align}
  c s_z^\prime + \dot{s} &= -g(s) s  + \gamma s_x  \label{eq_ds} \\
  c s^\prime + \dot{s}_z &= -g(s) s_z - \varepsilon s_y. \label{eq_dsz}
\end{align}

Another possible pair of variables $\Phi$ and $\phi$, the global and the relative phases of the two components, are defined by
\begin{equation}
\Phi = \frac{1}{4\mi} \mathrm{ln}\left( \psi_+ \psi_- \over \psi_+^* \psi_-^*\right), \qquad
\phi = \frac{1}{4\mi} \mathrm{ln}\left( \psi_+^* \psi_- \over \psi_+ \psi_-^*\right).
\end{equation}
This pair of variables, together with $s_z$ and $s$, fully defines the spinor components
\begin{equation} \label{eq_psi_vs_s}
\psi_\pm = \sqrt{s \pm s_z} \exp \left[ \mi\left(\Phi \mp \phi \right) \right].
\end{equation}
Eqs.\ (\ref{eq_ds}-\ref{eq_dsz}) may be supplemented with a pair of evolution equations on $\Phi$ and $\varphi$:
\begin{align}
c \phi^\prime - \dot{\Phi} = {1 \over 2} \left[ 3\alpha s + {\gamma s_z s_y - \varepsilon s s_x \over s_x^2 + s_y^2} \right], \label{eq_Phi} \\
c \Phi^\prime - \dot{\phi} = {1 \over 2} \left[ \alpha s_z + {\gamma s s_y - \varepsilon s_z s_x \over s_x^2 + s_y^2} \right], \label{eq_phi}
\end{align}
with
\begin{equation} \label{eq_phi_vs_sxy}
s_x = \sqrt{s^2 - s_z^2}\,\cos(2\phi), \;s_y = \sqrt{s^2-s_z^2}\,\sin(2\phi).
\end{equation}

We search for a plane wave solution
\begin{equation}\label{eq_cpmconst}
\psi_\pm(t) = \psi_\pm \exp(\mi Kx-\mi\Omega t),
\end{equation}
corresponding to $\mathbf{s}(x,t) = \textit{const}$.
Eqs.\ (\ref{eq_ds},\ref{eq_dsz}) then allow to express $s_x$ and $s_y$ components as
\begin{equation} \label{eq_sxy}
s_x = {g(s) \over \gamma} s, \qquad s_y = -{g(s) \over \varepsilon}  s_z.
\end{equation}
Taking into account that $s_x^2 + s_y^2 + s_z^2 = s^2$, we rewrite Eqs.\ \eqref{eq_sxy}, arriving at
\begin{align}
s_x &= \frac{sg(s)}{\gamma}, \label{eq_sx} \\
s_y &= \mp \frac{sg(s)}{\gamma} \sqrt{\frac{\gamma^2 - g(s)^2}{\ve^2 + g(s)^2}}, \label{eq_sy} \\
s_z &= \pm \frac{\ve s}{\gamma}\sqrt{\frac{\gamma^2 - g(s)^2}{\ve^2 + g(s)^2}}. \label{eq_sz}
\end{align}

In turn, Eqs.\ (\ref{eq_Phi},\ref{eq_phi}) allow to express the condensate energy $\Omega$ as a function of population $s$,
\begin{equation}\label{eq_Omega}
  \Omega =\frac{1}{2}\left[ 3 \alpha s - \frac{\varepsilon\gamma}{g(s)} \right],
\end{equation}
and find the equation for $s$:
\begin{equation}
\alpha s = \frac{\gamma\left(\varepsilon^2+g(s)^2\right)}{\ve g(s)} 
        \pm 2cK\frac{\gamma}{\varepsilon}\sqrt{\frac{\varepsilon^2+g(s)^2}{\gamma^2 - g(s)^2}} .  \label{eq_s}
\end{equation}
Eq.\ \eqref{eq_s} is not valid when $g(s) = \pm \gamma $. These two exclusions correspond to a pair of trivial solutions 
of Eqs.\ (\ref{eq_sx}-\ref{eq_sz}):
\begin{subequations}\label{eq_trivsol}
\begin{equation}
s=(W-\Gamma+\gamma)/\eta, \qquad s_x=s, \quad s_y = s_z = 0, \qquad (\mathrm{for}\;W>\Gamma-\gamma),
\end{equation}
\begin{equation}
s=(W-\Gamma-\gamma)/\eta, \qquad s_x=-s, \quad s_y = s_z = 0, \qquad (\mathrm{for}\;W>\Gamma+\gamma).
\end{equation}
\end{subequations}

We note that for $K=0$ we may derive the evolution equations on the other two pseudospin components $s_x$ and $s_y$, multiplying Eq.\ \eqref{eq_GP_2modes} by $\psi_\mp$ in the same manner as we get Eqs.\ (\ref{eq_psi1}-\ref{eq_dsz}):
\begin{equation} \label{eq_sxy_evol}
{\partial \over \partial t} \left(s_x + i s_y \right) = - \left[ g(s) + i \alpha s_z \right] \left(s_x + i s_y \right) + \gamma s + i \varepsilon s_z.
\end{equation}
After separating the real and imaginary parts in Eq.\ \eqref{eq_sxy_evol} and omitting the spatial derivative in Eq.\ \eqref{eq_dsz}, we obtain
\begin{align}
\dot{s_x} &= - g(s) s_x + \alpha s_z s_y + \gamma s, \nonumber \\
\dot{s_y} &= - g(s) s_y - \alpha s_z s_x + \varepsilon s_z,\\
\dot{s_z} &= - g(s) s_z - \varepsilon s_y, \nonumber
\end{align}
or, in the vectorial form:
\begin{equation} \label{eq_svect}
\dot{\mathbf{s}} = - g(s) \mathbf{s} + \alpha \left[ s_z \mathbf{e}_z \times \mathbf{s} \right] + (\gamma s, \ve s_z, -\ve s_y),
\end{equation}
where $\mathbf{e}_z$ is the $z$ axis unit vector.

In what follows we consider $K=0$. In this case, the population of nontrivial solutions is to be found from equation
\begin{equation} \label{eq_s_k0}
\alpha s = \frac{\gamma}{g(s)\varepsilon} \left[ \varepsilon^2 + g(s)^2 \right].
\end{equation}
We chose units so that $c = 1$ and normalize $\alpha$ and $\ve$ on $\eta$ and $\gamma$, respectively. The roots of the quadratic Eq.\ \eqref{eq_s_k0}
\begin{equation} \label{eq_s12}
s_{\pm} = \frac{(\beta - 2)w \pm\sqrt{\beta^2 w^2 + 4(\beta - 1)\varepsilon^2} }{2(\beta-1)},
\end{equation}
where $w = W-\Gamma$ is the relative pumping and $\beta = \alpha \ve$, satisfy the relations
\begin{align}
  s_+ + s_- =& \frac{\beta - 2}{\beta - 1} w, \\
  s_+ s_- =& - \frac{\ve^2 + w^2}{\beta - 1}.
\end{align}
Note that any physical (i.e., $s>0$) root of Eq.\ \eqref{eq_s_k0} corresponds to a pair of nontrivial fixed points of Eq.\ \eqref{eq_svect}, having opposite projections on the $z$ axis, and thus describing symmetry breaking polariton currents in opposite directions.

For the pair of trivial solutions, characterized by pseudospins oriented along $x$, the condensate population reads
\begin{equation}
s_S = w + 1, \qquad s_A = w - 1.
\end{equation}
The condition $\vert g(s) \vert < \vert \gamma \vert$ (see Eqs.\ (\ref{eq_sy},\ref{eq_sz})) implies that the nontrivial solutions are characterized by intermediate populations:
\begin{equation}
  w-1 \leqslant s_{\pm} \leqslant w+1.
\end{equation}
The trivial ($s_{S(A)}(w)$) and nontrivial ($s_{\pm}(w)$) branches intersect at the critical pumping
\begin{equation}
w_c = \text{sgn}(\beta) \left[ (1+ \ve^2)/\beta - 1 \right].
\end{equation}
If the determinant of the Eq.\ \eqref{eq_s_k0} is positive at some range of pumpings $w_f<w<w_c$, the nontrivial solutions are allowed for any $w > w_f$:
\begin{equation}
w_f = 2 \sqrt{ 1- \beta }/ \alpha.
\end{equation}
This type-I case is realized if the two characteristic parameters $\alpha$ and $\ve$, being the real to imaginary relations of the nonlinearity and the potential respectively, satisfy
\begin{equation}
\alpha < \ve^{-1} - \ve.
\end{equation}
Otherwise, the nontrivial solutions only appear for $w>w_c$.
This type-II case is, on the contrary, realized if $\alpha > \ve^{-1} - \ve$.

All possible configurations of solutions $s(w)$ are shown in Fig.\ S1.
We consider positive and negative $\ve$ separately, assuming positive $\alpha$ without reducing the generality, as the equations are only sensitive to the sign of $\beta$, or the relative sign of $\alpha$ and $\ve$.
In Fig.\ S1 we separate type-I and type-II cases for both signs of $\beta$.

\begin{figure}[h]
\includegraphics[scale=0.3]{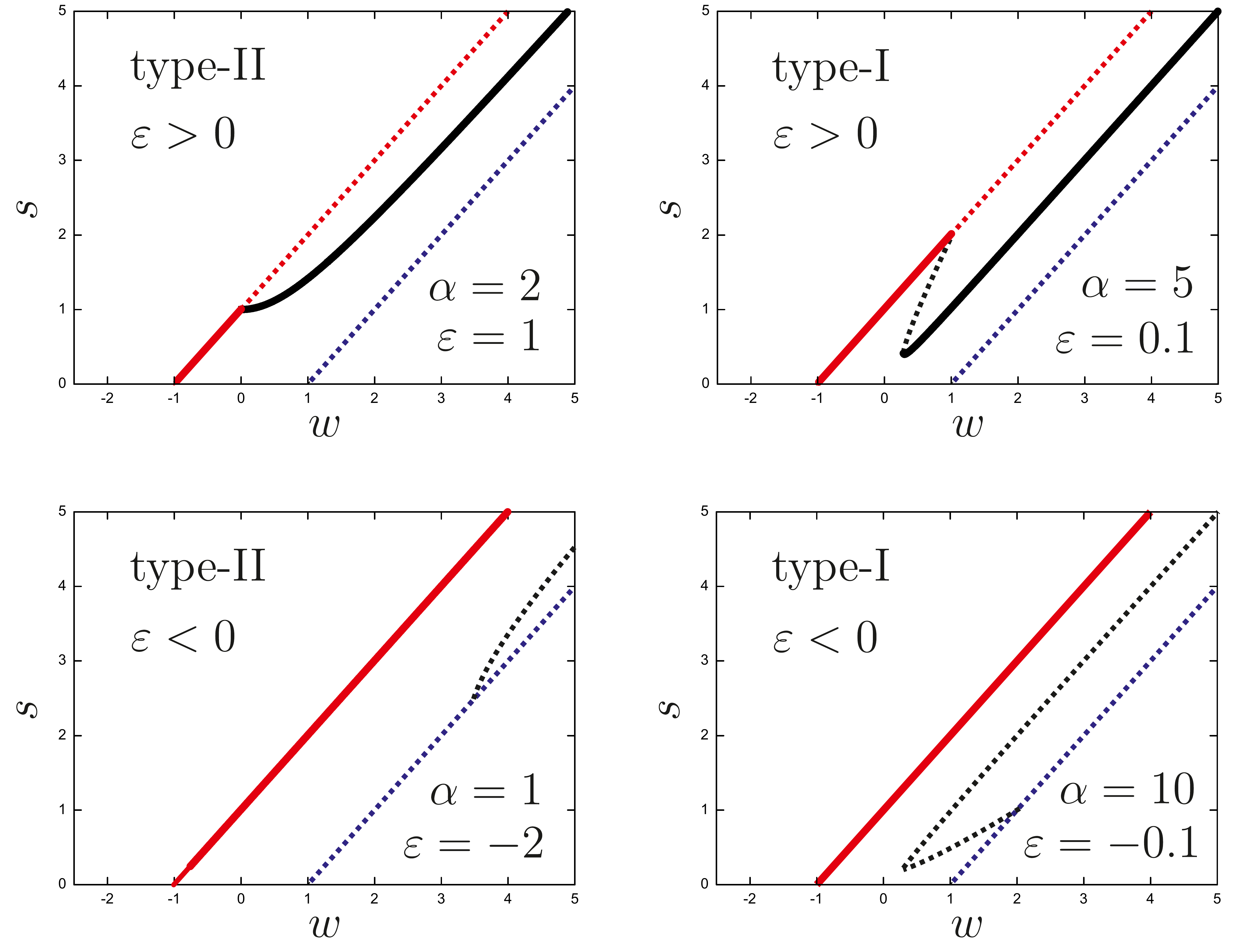}
\caption{Condensate population $s$ versus relative dimensionless pumping $w$. Red and blue lines correspond to the longest and the shortest lifetime single particle states at the opposite sides of the band gap. The black lines correspond to the symmetry breaking solutions. Solid(dashed) lines correspond to stable(unstable) stationary points of the pseudospin evolution equation. Two quantum phase transition types and two signs of $\beta$ are considered.}
\end{figure}

\section{Stability of the pseudospin evolution equation}

We study the stability of the symmetric and asymmetric solutions using the Jacobian of $\mathbf{f(s)} = d \mathbf{s}/dt$, where $\mathbf{f(s)}$ is the right-hand part of Eq. \eqref{eq_svect}:
\begin{equation} \label{eq_J}
J(\mathbf{f}) = \left( \begin{matrix}
w - s + \left(1 - s_x \right) s_x /s & \left( 1 - s_x \right) s_y /s + \alpha s_z & \left( 1 - s_x \right) s_z /s + \alpha s_y \\
- s_x s_y /s - \alpha s_z & w-s-s_y^2 /s & -s_y s_z /s + \varepsilon - \alpha s_x \\
- s_x s_z / s & - s_y s_z / s - \varepsilon & w-s-s_z^2 /s
\end{matrix} \right).
\end{equation}

Substituting trivial symmetric solutions $s = w \pm 1$, $s_x = 1 \pm w$, $s_y=s_z=0$ we obtain:
\begin{equation}
J_{tr} = \left( \begin{matrix}
-s & 0 & 0\\
0 & \mp 1 & \varepsilon \mp \alpha s\\
0 & - \varepsilon & \mp 1
\end{matrix} \right).
\end{equation}

All real parts of the Jacobian eigenvalues, or the Lyapunov exponents, have to be negative for a state to be stable.
While $-s$ is by definition negative, the other two eigenvalues are obtained from
\begin{equation}
(\lambda \pm 1)^2 + \ve^2 \mp \beta s = 0.
\end{equation}
For the lowest threshold trivial solution ($s = w + 1$) this yields
\begin{equation}
\lambda_\pm = -1 \pm \sqrt{\beta (w+1) - \ve^2}.
\end{equation}
Depending on the sign of $\beta$ this solution is either stable everywhere ($\beta < 0$), or stable for $w<w_c$ ($\beta > 0$).
The other trivial solution, characterized by Lyapunov exponents
\begin{equation}
\lambda_\pm = 1 \pm \sqrt{-\beta (w+1) - \ve^2},
\end{equation}
is obviously unstable.

Substitution of the nontrivial solutions \eqref{eq_s12} into Eq.\ \eqref{eq_J} leads to a somewhat cumbersome expression for the Jacobian:
\begin{equation}
J_\pm = \left( \begin{matrix}
w -s + \left[ 1-g(s) s\right] g(s) &
 \mp \left[ 1 - g(s) s \right] g(s) \sqrt{\gamma^2 - g(s)^2 \over \ve^2 + g(s)^2} &
  \pm \left[ 1 -g(s) s \right] \ve \sqrt{\gamma^2 - g(s)^2 \over \ve^2 + g(s)^2} \\
\pm s \left[ g(s)^2 - \beta \right] \sqrt{\gamma^2 - g(s)^2 \over \ve^2 + g(s)^2} &
w - s - s g(s)^2 {\gamma^2 - g(s)^2 \over \ve^2 + g(s)^2} &
s g(s) \ve {\gamma^2 - g(s)^2 \over \ve^2 + g(s)^2} + \ve - \alpha g(s) s \\
\mp s g(s) \ve \sqrt{\gamma^2 - g(s)^2 \over \ve^2 + g(s)^2} &
s g(s) \ve {\gamma^2 - g(s)^2 \over \ve^2 + g(s)^2} - \ve &
w -s - s \ve^2 {\gamma^2 - g(s)^2 \over \ve^2 + g(s)^2}
\end{matrix} \right).
\end{equation}
The eigenvalues of $J_\pm$ may change sign at the critical points $w=w_c$ and $w = w_f$ (in the type-I case), where the determinant $\vert J_\pm \vert = 0$.
The actual sign of the Lyapunov exponents may be traced with a simple numerical analysis.

Both trivial and symmetry breaking stable solutions are shown in Fig.\ S1 by solid lines, while unstable solutions are shown by dotted lines.

The case $\beta < 0$ is trivial in terms of stability: nontrivial solutions are always unstable, the only stable solution is characterized with the lowest threshold.
If $\beta > 0$ the lower nontrivial branch $s_-(w)$ is stable for $w>w_c$ in the type-II case and for $w>w_f$ in the type-I case.

\section{Nonlocal stability}

The Lyapunov exponents of Eq.\ \eqref{eq_svect} characterize the stability of the stationary solutions with respect to spatially homogeneous fluctuations.
To study the general case of inhomogeneous fluctuations, characterized with a wavevector $k\ne0$, one can write the uniform ($K=0$) solutions \eqref{eq_psi_vs_s} as 
\begin{equation}\label{eq_ns1}
\psi_\pm=e^{\mi\Phi}z_\pm, \qquad \Phi=-\Omega t, \qquad z_\pm=\sqrt{s\pm s_z}\,e^{\mp\mi\phi},
\end{equation}
where the frequency $\Omega$, the asimuthal angle $\phi=\half\tan^{-1}(s_y/s_x)$, and the pseudospin components are given by Eqs.\ (\ref{eq_sx}--\ref{eq_Omega}) and \eqref{eq_s_k0}. It follows from Eq.\ \eqref{eq_GP_2modes} that $z_\pm$ satisfy 
\begin{equation}
  -\mi\Omega z_\pm = \frac{1}{2}\left[ w - \frac{\eta}{2}(\vert z_+\vert^2 + \vert z_-\vert^2 ) 
                                      - \mi\alpha (\vert z_\pm \vert^2 + 2 \vert z_\mp\vert^2) \right] z_\pm 
                                      +\frac{1}{2}(\gamma+\mi\varepsilon)z_\mp.
\end{equation}
Then we substitute
\begin{equation}
\psi_\pm(x,t) = e^{-\mi\Omega t}\left[z_\pm + u_\pm e^{\mi kx + \lambda t} + v^*_\pm e^{-\mi kx + \lambda^* t}\right],
\end{equation}
into Eq.\ \eqref{eq_GP_2modes} and keep the terms linear in the amplitudes $u_\pm$ and $v_\pm$. As a result, we obtain the eigenvalue equations, which in matrix form are
\begin{equation} \label{eq_excitation_matrix1}
 \left( \begin{matrix}
  M_+ -\mi ck  &  L_+  &  P_+  &  Q  \\
  L_+^*  &  M_+^* -\mi ck  &  Q^*  &  P_+^*  \\
  P_- &  Q  &  M_- +\mi ck  &  L_- \\
  Q^*  &  P_-^*  &  L_-^*  &  M_-^* +\mi ck  
 \end{matrix} \right) 
\left( \begin{matrix} u_+ \\ v_+ \\ u_- \\ v_- \end{matrix} \right)
 =\lambda \left( \begin{matrix} u_+ \\ v_+ \\ u_- \\ v_- \end{matrix} \right),
\end{equation}
or
\begin{equation} \label{eq_excitation_matrix2}
 \left( \begin{matrix}
  M_+ -\mi ck  &  P_+  &  L_+  &  Q  \\
  P_-  &  M_- +\mi ck  &  Q  &  L_-  \\
  L_+^*  &  Q^*  &  M_+^* -\mi ck  &  P_+^*  \\
  Q^*  &  L_-^*  &  P_-^*  &  M_-^* +\mi ck  
 \end{matrix} \right) 
\left( \begin{matrix} u_+ \\ u_- \\ v_+ \\ v_- \end{matrix} \right)
 =\lambda \left( \begin{matrix} u_+ \\ u_- \\ v_+ \\ v_- \end{matrix} \right),
\end{equation}
where
\begin{align}\label{eq_matpar}
  & M_\pm = \frac{w}{2} + \mi\Omega - \mi\Big(\alpha-\frac{\mi\eta}{2}\Big)|z_\pm|^2 - \mi\Big(\alpha-\frac{\mi\eta}{4}\Big)|z_\mp|^2, \\
  & L_\pm = -\frac{\mi}{2}\Big(\alpha-\frac{\mi\eta}{2}\Big)z_\pm^2, \\
  & P_\pm = \frac{1}{2}(\gamma+\mi\ve)-\mi\Big(\alpha-\frac{\mi\eta}{4}\Big)z_\pm z_\mp^*, \\
  & Q=-\mi\Big(\alpha-\frac{\mi\eta}{4}\Big)z_+ z_-.
\end{align}
The results of the numerical analysis of the above system of equations are given in the main text.

\section{Discussion of experimental realization}

The spontaneous current phase can be conveniently identified in the interferometry experiments that are used to access the phase maps of polariton condensates \cite{Sanvitto2010,Lagoudakis2009,Lagoudakis2011,Nardin2010}.
Moreover, one can find a signature of a polariton domain in the angular distribution of the far field emission of the microcavity which is expected to be asymmetric in the case of the spatial inversion symmetry breaking.

We estimate the characteristic scales of the effect, in particular, the domain size and domain wall velocity, for a structure similar to the one studied in Ref.\ \cite{Lai2007a}.
The complex potential for polaritons is realized with metallic grating deposited on the top of the microcavity and is characterized with period of the order of several micrometers.
The real part of the potential modulation of the order of $\varepsilon \sim 100\,\mu$eV stems from the cavity mode blueshift in the regions of microcavity under the metal cover.
Neglecting the quality factor inhomogeneity, the decay rate modulation, corresponding to the imaginary part of the potential, stems from the spatial variation of the Hopfield coeffient, or the excitonic fraction of polariton eigestates at the lower polariton branch.
Its value may be majorated by the photon escape rate from the cavity, which is typically $\gamma \sim (10\,$ps$)^{-1}$.
Since the characteristic domain wall velocity value in this structure is $v_d \approx c = (\pi \hbar)/(m a) \sim 10^9\,$cm/s, the minimal domain size in the vicinity of the bifurcation point may be approximated by $l_d \sim c/\gamma \sim 100\,\mu$m, which exceeds the excitation spot size used in Ref. \cite{Lai2007a}.
One may therefore expect angularly asymmetric emission from the polariton condensate, which then transforms into a dynamic domain structure at higher pumping powers.

The polariton repulsion constant may be estimated as \cite{Glazov2009} $\alpha = 6 \vert X \vert^2 R_y a_B^2/l$, where $X$ is the Hopfield coeffient, $R_y$ and $a_B$ are the exciton Rydberg energy and Bohr radius respectively, and $l$ is the confinement length in the lateral direction normal to the modulation axis.
The reservoir depletion constant $\eta$ is typically lower than the interaction constant $\alpha$ \cite{Keeling2008}.
This may be witnessed from the fact that polariton energy broadenings in optical traps are much lower than the confinement energies produced by the reservoir \cite{Askitopoulos2015a}.
The relation $ \alpha / \eta \gg 1$ combined with $\varepsilon / \gamma > 1$  means that one should expect type-II bifurcations towards stable spontaneous current phase.

\end{document}